\documentclass[preprintnumbers]{revtex4}
\usepackage{amsfonts}
\usepackage{amsmath}
\usepackage{graphicx}
\usepackage{dcolumn}
\usepackage{bm}

\input{tcilatex}

\begin{document}


\title{Two approaches in the theory of atom interferometry}
\author{ B. Dubetsky}
\affiliation{bdubetsky@gmail.com
}

 \date{\today}

\begin{abstract}

\end{abstract}

\pacs{03.75.Dg, 37.25.+k, 04.80.-y}
\maketitle

In the theory of atom interferometry (AI) \cite{c1}, one can choose between
several approaches. I use the approach based on the density matrix equation
in the Wigner representation (DMEWR) \cite{c2}. Starting from the article 
\cite{c3}, this is a traditional approach in the theory of phenomena related
to the quantization of the atomic center mass motion and laser cooling \cite%
{c4}. However, to my knowledge, it has been used sparingly in the context of
AI \cite{c5,c6}. The convenience of this approach is that, for the time
between Raman pulses, the density matrix obeys an equation that is similar
to the classical Liouville equation for the distribution function \cite{c2}.

Another approach here is path integrals (PI) \cite{c7}, which is used
starting from article \cite{c8}.

I am here going to compare these 2 approaches to a specific problem, the
part of the AI\ phase caused by an external test mass. The expression for
this part is used in measurements of the Newtonian gravitational constant $G$
using AI. This is a promising technique. Using a recently elaborated \cite%
{c22} error model, I showed that at the current state of the art in the AI 
\cite{c23,c24,c25,c26}, this measurement should be off 200ppb accuracy,
which is more than 2 orders of magnitude better than the accuracy accepted
now \cite{c27}. I justified the expression for the phase by applying DMEWR
approach \cite{c9}. While in real experiments \cite{c11,c12} the PI approach
was used. The PI approach has also been used in article \cite{c29}.

I am going to analytically compare these two approaches here. Consider atoms
moving in the gravity potential%
\begin{equation}
V\left( \vec{x}\right) =-M_{a}\vec{g}\cdot \vec{x}+\delta V\left( \vec{x}%
\right) ,  \label{1}
\end{equation}%
where $M_{a}$ is the atomic mass, $\vec{g}$ is Earth's gravity field, $%
\delta V\left( \vec{x}\right) $ is the gravity potential of the external
test mass, which we assume to be small,%
\begin{equation}
\delta V\left( \vec{x}\right) \ll V\left( \vec{x}\right) .  \label{2}
\end{equation}%
After interacting with $\pi /2-\pi -\pi /2$ sequence of Raman fields, the
atomic levels' populations acquire interferometric terms whose phase
contains the addition $\delta \phi ,$ caused by the external test mass
gravity field. Using DMEWR approach, we calculated \cite{c9} $\delta \phi $
for the arbitrarily moving test mass. In the case of the stationary test
mass, which we consider here, from Eqs. (58a,62,64,73,88) in article \cite%
{c9}, one finds 
\begin{subequations}
\label{3}
\begin{eqnarray}
\delta \phi  &=&\vec{k}\cdot \int_{0}^{T}dt\left\{ t\delta \vec{g}\left[ 
\vec{s}\left( t_{1}+t\right) +\dfrac{\hbar \vec{k}}{2M_{a}}t\right] +\left(
T-t\right) \delta \vec{g}\left[ \vec{s}\left( t_{1}+T+t\right) +\dfrac{\hbar 
\vec{k}}{2M_{a}}\left( T+t\right) \right] \right\} +\phi _{Q},  \label{3a} \\
\phi _{Q} &\approx &\dfrac{\hbar ^{2}}{24M_{a}^{2}}k_{i}k_{j}k_{l}%
\int_{0}^{T}dt\left\{ t^{3}\chi _{ijl}\left[ \vec{s}\left( t_{1}+t\right) %
\right] +\left( T-t\right) ^{3}\chi _{ijl}\left[ \vec{s}\left(
t_{1}+T+t\right) \right] \right\} ,  \label{3b} \\
\vec{s}\left( t\right)  &=&\vec{v}t+\QDABOVE{1pt}{1}{2}\vec{g}t^{2},
\label{3c}
\end{eqnarray}%
where $t_{1}$ is the time delay between the time of the atoms' launching $%
\left( t=0\right) $ and the 1st Raman pulse, $T$ is the time separation
between Raman pulses, 
\end{subequations}
\begin{subequations}
\label{4}
\begin{eqnarray}
\delta \vec{g}\left( \vec{x}\right)  &=&-\QDABOVE{1pt}{1}{M_{a}}\partial _{%
\vec{x}}\delta V\left( \vec{x}\right) ,  \label{4a} \\
\chi _{ijl}\left( \vec{x}\right)  &=&\partial _{x_{j}}\partial
_{x_{l}}\delta g_{i}\left( \vec{x}\right)   \label{4b}
\end{eqnarray}%
are the acceleration and the curvature tensor of the test mass field. In
Eqs. (\ref{3}) we assumed, for simplicity, that the atoms are launched from
the frame origin. One can expand $\delta \phi $ into a power series over the
Planck constant. Since only the 1st nonzero part of Q-term (\ref{3b}) has
been obtained in \cite{c9}, one should truncate the series for $\delta \phi $
with a second order term. Using that 
\end{subequations}
\begin{equation}
\delta g_{i}\left( \vec{x}+\vec{\varepsilon}\right) \approx \delta
g_{i}\left( \vec{x}\right) +\underline{\gamma }_{ij}\left( \vec{x}\right)
\varepsilon _{j}+\QDABOVE{1pt}{1}{2}\chi _{ijl}\left( \vec{x}\right)
\varepsilon _{j}\varepsilon _{l},  \label{5}
\end{equation}%
where%
\begin{equation}
\underline{\gamma }\left( \vec{x}\right) =\partial _{\vec{x}}\delta \vec{g}%
^{T}\left( \vec{x}\right)   \label{6}
\end{equation}%
is the gravity-gradient tensor, and the equality 
\begin{equation}
T^{3}-t^{3}=\QDABOVE{1pt}{1}{4}\left[ \left( T-t\right) ^{3}+3\left(
T-t\right) \left( T+t\right) ^{2}\right] ,  \label{7}
\end{equation}%
one obtains the following series%
\begin{eqnarray}
\delta \phi  &=&\vec{k}\cdot \int_{0}^{T}dt\left\{ t\delta \vec{g}\left[ 
\vec{s}\left( t_{1}+t\right) \right] +\left( T-t\right) \delta \vec{g}\left[ 
\vec{s}\left( t_{1}+T+t\right) \right] \right\}   \notag \\
&&+\dfrac{\hbar }{2M_{a}}\vec{k}\cdot \int_{0}^{T}dt\left\{ t^{2}\underline{%
\gamma }\left[ \vec{s}\left( t_{1}+t\right) \right] +\left(
T^{2}-t^{2}\right) \underline{\gamma }\left[ \vec{s}\left( t_{1}+T+t\right) %
\right] \right\} \vec{k}  \notag \\
&&+\dfrac{\hbar ^{2}}{6M_{a}^{2}}k_{i}k_{j}k_{l}\int_{0}^{T}dt\left\{
t^{3}\chi _{ijl}\left[ \vec{s}\left( t_{1}+t\right) \right] +\left(
T^{3}-t^{3}\right) \chi _{ijl}\left[ \vec{s}\left( t_{1}+T+t\right) \right]
\right\} .  \label{8}
\end{eqnarray}

\begin{figure}[!t]
\includegraphics[width=13cm]{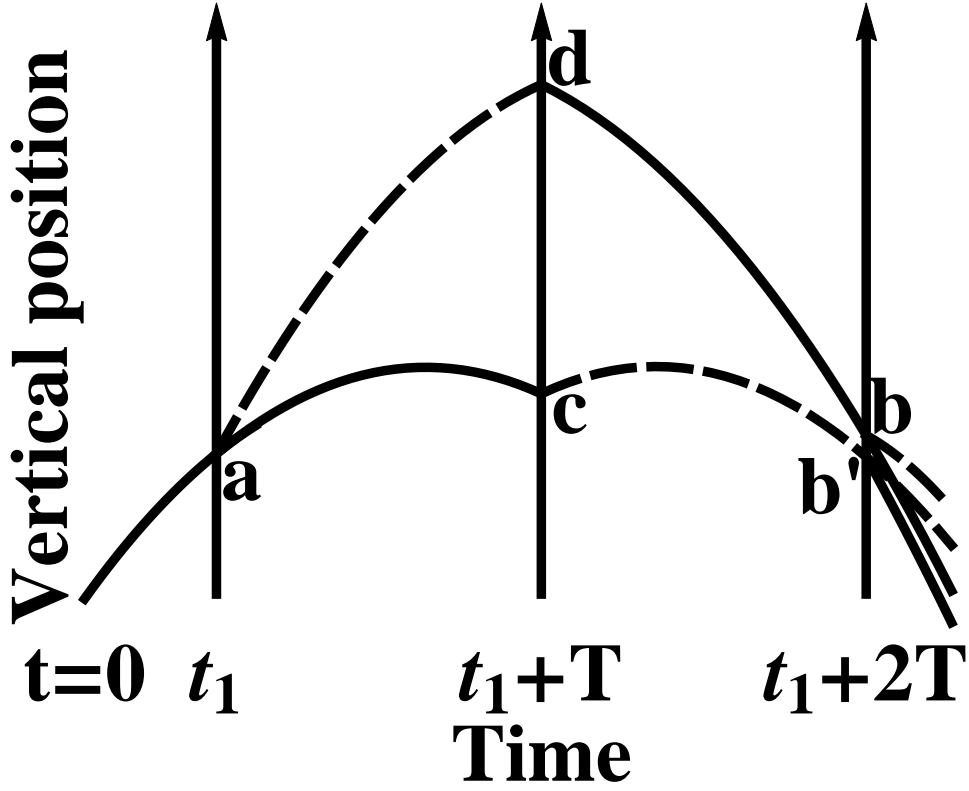}
  \caption{Atomic trajectories in the gravity field (\protect\ref{1}) after
interacting with three Raman pulses at moments $\left\{
t_{1},t_{1}+T,t_{1}+2T\right\} .$ Solid and dashed curves correspond to the
ground and excited atomic states.}
\label{f}
\end{figure}

Consider now the PI approach. In this approach, one can calculate the phase
in 2 different ways \cite{c30,c34}, using the expression%
\begin{equation}
\phi =\phi _{prop}+\phi _{laser}+\phi _{sep},  \label{9}
\end{equation}%
where $\phi _{prop},~\phi _{laser},~\phi _{sep}$ are the so-called
propagation, laser and separation \cite{c34} phases, or using an approximate
expression for the addition to phase \cite{c30}, valid at condition (\ref{2}%
) 
\begin{equation}
\delta \phi =-\QDABOVE{1pt}{1}{\hbar }\doint\limits_{adbca}dt\delta V\left[ 
\vec{x}_{0}\left( t\right) \right] ,  \label{10}
\end{equation}%
where $\vec{x}_{0}\left( t\right) $ is the atom trajectory in the absence of
test-mass potential (see Fig. \ref{f}).

Expression (\ref{9}) was used in experiment \cite{c11}, while Eq. (\ref{10})
was used in experiment \cite{c12} and article \cite{c29}. The equivalence of
the Eqs. (\ref{9}) and (\ref{10}) has been verified for the linear gravity
potential \cite{c30}. After making some calculations, I also verified this
equivalence for arbitrary small nonlinear stationary test mass potential 
\cite{c35}.

For trajectories shown in Fig. \ref{f}, one gets 
\begin{equation}
\left\{ 
\begin{array}{c}
\vec{x}_{0}\left( t\right) _{ad} \\ 
\vec{x}_{0}\left( t\right) _{ac} \\ 
\vec{x}_{0}\left( t\right) _{db} \\ 
\vec{x}_{0}\left( t\right) _{cb^{\prime }}%
\end{array}%
\right\} =\vec{s}\left( t\right) +\QDABOVE{1pt}{\hbar \vec{k}}{M_{a}}\left( 
\begin{array}{c}
t-t_{1} \\ 
0 \\ 
T \\ 
t-t_{1}-T%
\end{array}%
\right) ,  \label{11.}
\end{equation}%
where $\vec{s}\left( t\right) $ is given by Eq. (\ref{3c}). Using these
trajectories, one can represent the contour integral in Eq. (\ref{10}) as 2
integrals from $t_{1}$ to $t_{1}+T$ and from $t_{1}+T$ to $t_{1}+2T$.
Transforming both integrals to those from zero to $T$, one finds

\begin{equation}
\delta \phi =-\QDABOVE{1pt}{1}{\hbar }\int_{0}^{T}dt\left\{ \delta V\left[ 
\vec{s}\left( t_{1}+t\right) +\dfrac{\hbar \vec{k}}{M_{a}}t\right] -\delta V%
\left[ \vec{s}\left( t_{1}+t\right) \right] +\delta V\left[ \vec{s}\left(
t_{1}+T+t\right) +\dfrac{\hbar \vec{k}}{M_{a}}T\right] -\delta V\left[ \vec{s%
}\left( t_{1}+T+t\right) +\dfrac{\hbar \vec{k}}{M_{a}}t\right] \right\} .
\label{12}
\end{equation}%
Finally, using for potential expansion%
\begin{equation}
\delta V\left( \vec{x}+\vec{\varepsilon}\right) \approx \delta V\left( \vec{x%
}\right) -M_{a}\left[ \delta \vec{g}\left( \vec{x}\right) \vec{\varepsilon}+%
\QDABOVE{1pt}{1}{2}\vec{\varepsilon}\cdot \underline{\gamma }\left( \vec{x}%
\right) \vec{\varepsilon}+\QDABOVE{1pt}{1}{6}\chi _{ijl}\left( \vec{x}%
\right) \varepsilon _{i}\varepsilon _{j}\varepsilon _{l}\right]  \label{13}
\end{equation}%
one arrives at the series (\ref{8}).

The coincidence of the power series for the phase $\delta \phi $ in DMEWR-
PI-approaches means that Q-term \cite{c9} has to be included in
DMEWR-approach, while no Q-term arises in PI-approach.

\textbf{Acknowledgments}

I would like to thank Dr. M. Kasevich who suggested the consideration of
this problem from me. I am also grateful to Dr. M. Prevedelli for
discussions and the numerical verification of the Eqs. (\ref{3}, \ref{10})
equivalence, and to Dr. A. Landragin for discussions.

\end{document}